\documentclass[aps,pre,floatfix,twocolumn,showpacs,nofootinbib,superscriptaddress]{revtex4-1}
\usepackage{amssymb}
\usepackage{amsmath}
\usepackage{amsfonts}
\usepackage{epsfig}
\usepackage{graphicx}
\usepackage{color}
\usepackage[latin1]{inputenc}
\usepackage{hyperref}

\begin{document}

\author{Liudmila Rozanova}
\affiliation{Departament de F\'isica de la Mat\`eria Condensada, Universitat de Barcelona, Mart\'i i Franqu\`es 1, 08028 Barcelona, Spain}
\email{rozanova@ffn.ub.edu}
\author{Mari\'an Bogu\~{n}\'a}
\affiliation{Departament de F\'isica de la Mat\`eria Condensada, Universitat de Barcelona, 
Mart\'i i Franqu\`es 1, 08028 Barcelona, Spain}
\affiliation{Universitat de Barcelona Institute of Complex Systems (UBICS), Universitat de Barcelona, Barcelona, Spain}
\email{marian.boguna@ub.edu}
\date{\today}

\title{Dynamical properties of the herding voter model with and without noise}

\begin{abstract}
Collective leadership and herding may arise in standard models of opinion dynamics as an interplay of a strong separation of time scales within the population and its hierarchical organization. Using the voter model as a simple opinion formation model, we show that, in the herding phase, a group of agents become effectively the leaders of the dynamics while the rest of the population follow blindly their opinion. Interestingly, in some cases such herding dynamics accelerates the time to consensus, which then become size independent or, on the contrary, makes the consensus nearly impossible. These new behaviors have important consequences when an external noise is added to the system that makes consensus (absorbing) states to disappear. We analyze this new model which shows an interesting phase diagram, with a purely diffusive phase, a herding (or two-states) phase, and mixed phases where both behaviors are possible.
\end{abstract}


\maketitle

\section{Introduction}
The study of the behavior of interacting populations, being these human populations, groups of cells within organs, or colonies of social insects, is a key element to understand, predict, and control their function at the global scale. Despite the stochastic nature of individual agents within the population, the global dynamics of the system can, in many cases, be analyzed using tools, models, and techniques from statistical physics~\cite{Arnopoulos:1993,Galam:2012,Sen_Book,Castellano:2008ge}. A good example is opinion dynamics, that is, the study of the rules that govern transitions between different opinion states as a response to social influence --the tendency of people to behave like their peers-- and the effects that such rules have on the global opinion state of the population~\cite{Festinger:1950}. The simplest model with the simplest rules is the voter model, which is the object of our present study.

The voter model is one of the most paradigmatic and popular models of opinion dynamics~\cite{Clifford:1973zc,Holley:1975}. It has been used to model different phenomena in both the natural and social sciences, from catalytic reaction models~\cite{EVANS:1991kx,Evans:1993fk} to the evolution of bilingualism~\cite{Castello:2006ik} or US presidential elections~\cite{Fernandez-Gracia:2014uq}. The original voter model is defined as follows. There is a set of $N$ interacting agents, each endowed with a binary state of opinion (sell or buy, Democrat or Republican, Windows or Mac, etc). At each time step of the simulation, an agent is randomly chosen to interact with one of her social contacts, after which the agent copies the opinion of her contact. The model has two absorbing (frozen) states, which correspond to the two consensus states, and one important property is then the average consensus time, that is, the average time it takes the dynamic to reach consensus starting from a given initial configuration. In the voter model, the noise term associated to the global behavior of the system typically decreases with the system size and, as a consequence, the average consensus time is a growing function of the size of the system. The specific size-dependence is directly related to the pattern of interactions among agents~\cite{Sood:2005fk,Suchecki:2005fk,Sood:2008kk}.

One important question in opinion dynamics models is their ability to generate spontaneously collective leadership, that is, a group of agents that, spontaneously would agree in their opinion whereas the rest of the population would follow such opinion blindly. This phenomenon is most probably related to the strong fluctuations observed in stock markets during speculative periods or severe crisis. While such phenomenon cannot arise in standard voter models, in~\cite{Mosquera-Donate:2015aa}, we introduced a novel generalization, the herding voter model, which is able to show emergence of collective leadership as a response to strong heterogeneity in the activity patterns of agents and a structured influence matrix. In this paper, we study the dynamical properties of the herding voter model with and without intrinsic noise. The introduction of intrinsic noise in any variation of the voter model has the effect of removing the frozen states and making the steady state possible. As we shall see below, when coupled to the herding voter model, it has quite interesting and unexpected consequences. As for the herding voter model without noise, we study analytically the average consensus time, showing that, depending on the model parameters, it may range from being size independent to have an exponential dependence on the system size, turning then the absorbing states unreachable.

\section{The herding voter model}

The herding voter model introduced in~\cite{Mosquera-Donate:2015aa} takes into account simultaneously heterogeneous populations where agents are given intrinsic activity rates $\{ \lambda_i\}$~\cite{Masuda:2010uq,Fernandez-Gracia:2011kx} ---accounting for the rate at which agents interact with their peers--- and arbitrary influences of agents on each other. This is modeled through the probability $\mathrm{Prob}(j|i)$ that agent $i$ copies the opinion of agent $j$ when $i$ is activated at rate $\lambda_i$. In the herding voter model, this probability is taken to be a function of the activity rate of the copied agent of the form
\begin{equation}
	\mathrm{Prob}(j|i)=\frac{f(\lambda_{j})}{\sum_{i=1}^N f(\lambda_{i})},
	\label{eq:Pij}
\end{equation}
where $f(\lambda_{j})$ is an arbitrary function. When $f(\lambda_{j})$ is a monotonic increasing function, active agents are chosen more frequently and, in the herding phase, their opinions have a strong influence on the entire population. To simplify the analysis, hereafter we consider a structured population with only two activity rates, $\lambda_s \ll \lambda_f$, corresponding to $N_s$ slow and $N_f$ fast agents, respectively.

The dynamics of the state of the system can be described using a set of $N=N_f+N_s$ dichotomous stochastic processes $\{n_i(t)\}$ taking the value $0$ or $1$ depending on the opinion state of each agent at time $t$. The homogeneity within each segment of the population allows us to coarse-grain the system by defining the instantaneous average state of each sub-population as 
\begin{equation}
\Gamma_f(t) \equiv \frac{1}{N_f}\sum_{i\in fast} n_i(t) \; ; \; \Gamma_s(t) \equiv \frac{1}{N_s}\sum_{i\in slow} n_i(t),
\end{equation} where $i\in fast$ ($i\in slow$) means summation over all fast (slow) agents.
In the limit of large system sizes, $\Gamma_f(t)$ and $\Gamma_s(t)$ can be considered as quasi-continuous stochastic processes evolving in the range $[0,1]$ that can be described by a Langevin equation. In particular, for the dynamics of the fast group, we can write:
\begin{equation}
	\frac{d\Gamma_f(t)}{dt}=A_f\left[\vec{\Gamma}(t)\right]+\sqrt{D_f\left[\vec{\Gamma}(t)\right]}\xi_f(t),
\label{eq:langevin}
\end{equation} 
where $\xi_f(t)$ is Gaussian white noise. Both the drift and diffusion terms appearing in this equation where computed exactly in~\cite{Mosquera-Donate:2015aa} and read:
\begin{equation}
A_f=\alpha_{fs}(\Gamma_{s}-\Gamma_{f})
\label{eq:drift}
\end{equation}      
\begin{equation}
D_f=\frac{\alpha_{fs}}{N_f} \left( \Gamma_s+\Gamma_f \left[ 1+2 \beta_{fs}-2 \Gamma_s -2 \beta_{fs} \Gamma_f\right] \right),
\label{eq:diffusion}
\end{equation}
where we have defined: 
\begin{equation}
\alpha_{fs}=\frac{\lambda_f}{1+\beta_{fs}} \; \mbox{ and } \; \beta_{fs}=\frac{N_f f(\lambda_{f})}{N_s f(\lambda_{s})}. 
\end{equation}
Similar equations can be derived for the slow group by replacing the index $f \leftrightarrow s$ in the preceding equations.

The main finding of~\cite{Mosquera-Donate:2015aa} is the discovery of a phase transition between a diffusive phase and a herding phase whenever the following condition is met
\begin{equation}
2\frac{f(\lambda_f)}{f(\lambda_s)}>N_s.
\label{eq:critical}
\end{equation}
In the herding phase, the group of fast agents behaves as a two-states system such that their aggregated opinion fluctuates close to zero during some random time, then jumps quickly to one, remains there for another random time, and jumps back to zero again and so on. During the periods when fast agents have a stable opinion around zero or one, the group of slow agents follow quasi-deterministically the opinion of the fast group. The intuitive explanation of this phenomenon is as follows. When the separation of time scales is large ({\it i.~e.} $\lambda_f \gg \lambda_s$) fast agents perceive slow ones as frozen in a given state and these will act as a drift for the evolution of fast agents. When the condition Eq.~\eqref{eq:critical} is satisfied, this drift is small enough so that fast agents evolve almost freely until they reach one of the consensus states (either zero or one). In the absence of slow agents, the consensus state would be an absorbing state. However, when slow agents have an average opinion different from zero or one, they eventually take fast agents out of the consensus state and make them to jump to the other consensus state.

\subsection{Consensus time in the herding voter model}

Despite the separation of time scales, fast agents can spend a very long time near one of the consensus states. In this case, slow agents approach almost deterministically the state of fast agents with a characteristic time $\lambda_s^{-1}$. If the time needed by fast agents to switch globally their opinions is comparable to $\lambda_s^{-1}$ slow agents can reach the opinion of fast agents before the switching event takes place, making the change of opinion of fast agents even more difficult and increasing the probability that the entire system (fast and slow agents) reach the consensus ---and so frozen--- state. Therefore, the understanding of the global consensus time is necessarily related to the understanding of the first passage time from one of the boundaries to the other for fast agents when slow agents have a given state $\Gamma_s$.

We are interested in the herding phase in the thermodynamic limit $N_s>N_f \gg 1$. Following Eq.~\eqref{eq:critical}, we define the order parameter
\begin{equation}
x \equiv \frac{2 f(\lambda_f)}{N_s f(\lambda_s)}.
\end{equation} 
In the large size limit, and with $x$ constant, the Langevin equation describing the opinion of the fast group can be written as
\begin{equation}
\frac{d\Gamma_f(t)}{dt}=\frac{2\lambda_f}{x N_f}[\Gamma_s-\Gamma_f]+\sqrt{\frac{2 \lambda_f}{N_f}\Gamma_f(1-
\Gamma_f)}\xi_f(t),
\label{Langevin_fast}
\end{equation}
whereas for the slow group we have
\begin{equation}
\frac{d\Gamma_s(t)}{dt}=\lambda_s[\Gamma_f-\Gamma_s].
\end{equation}

Let us now suppose that the state of slow agents is fixed at some value $\Gamma_s$. We are interested in the average time it takes for fast agents to reach the boundary at $\Gamma_f=1-\Delta \Gamma$ starting from $\Gamma_f=\Delta \Gamma$. This is just the standard mean first passage time for a stochastic process following the Langevin equation Eq.~\eqref{Langevin_fast} with a reflecting boundary at $\Gamma_f=\Delta \Gamma$ and an absorbing one at $\Gamma_f=1-\Delta \Gamma$. Notice that, due to the discrete nature of the process, we take $\Delta \Gamma=\mathcal{O}(N_f^{-1})$. The solution can be written as~\cite{Gardiner:2004uv}
\begin{equation}
T_{f}=\frac{N_f}{\lambda_f} \int_{0}^{1-\Delta \Gamma} \frac{B(z,\frac{2\Gamma_s}{x},\frac{2(1-\Gamma_s)}{x})}{z^{\frac{2\Gamma_s}{x}}(1-z)^{\frac{2(1-\Gamma_s)}{x}}},
\end{equation}
where $B(z,a,b)$ is the incomplete Beta function. To get further insights, we chose $f(\lambda)=\lambda^{\sigma}$ and $N_f=a N_s^{\beta}$ with $\beta \le 1$. With this particular choice, the mean first passage time for the fast group becomes
\begin{equation}
\lambda_s T_{f}=a\left(\frac{2}{x}\right)^{1/\sigma} N_s^{\beta-1/\sigma} \int_{0}^{1-\Delta \Gamma} \frac{B(z,\frac{2\Gamma_s}{x},\frac{2(1-\Gamma_s)}{x})}{z^{\frac{2\Gamma_s}{x}}(1-z)^{\frac{2(1-\Gamma_s)}{x}}}dz.
\label{mfpt}
\end{equation}
The term $\lambda_s T_{f}$ is a dimensionless quantity which value determines the behavior of the global consensus time. 

When $\lambda_s T_{f} \gg1$ slow agents have enough time to decay to the same state as the fast group and, therefore, the global consensus time is $T_{con} \sim \lambda_s^{-1}$, independent of the size of the system. Instead, when $\lambda_s T_{f} \ll 1$ the fast group oscillates very rapidly between zero and one and the slow group does not have enough time to decay. In this case, we can estimate the global consensus time using an argument from extreme value theory. In general, consensus will be achieved when one of the crossing times is of the order of $\lambda_s^{-1}$. On average, the number of attempts before such event takes place at least once is proportional to $T_{con}/T_f$. If we assume that crossing times are exponentially distributed with average $T_f$, then we can write $T_{con}/T_f e^{-1/\lambda_s T_f} \sim 1$ and, thus the global consensus time scales as
\begin{equation}
T_{con} \sim T_f e^{1/\lambda_s T_f}.
\end{equation}  
When $T_f$ decays with the system size, the exponential dependence of $T_{con}$ on $T_f$ will make global consensus virtually impossible.
\begin{figure}[t]
\centering
\includegraphics[width=\columnwidth]{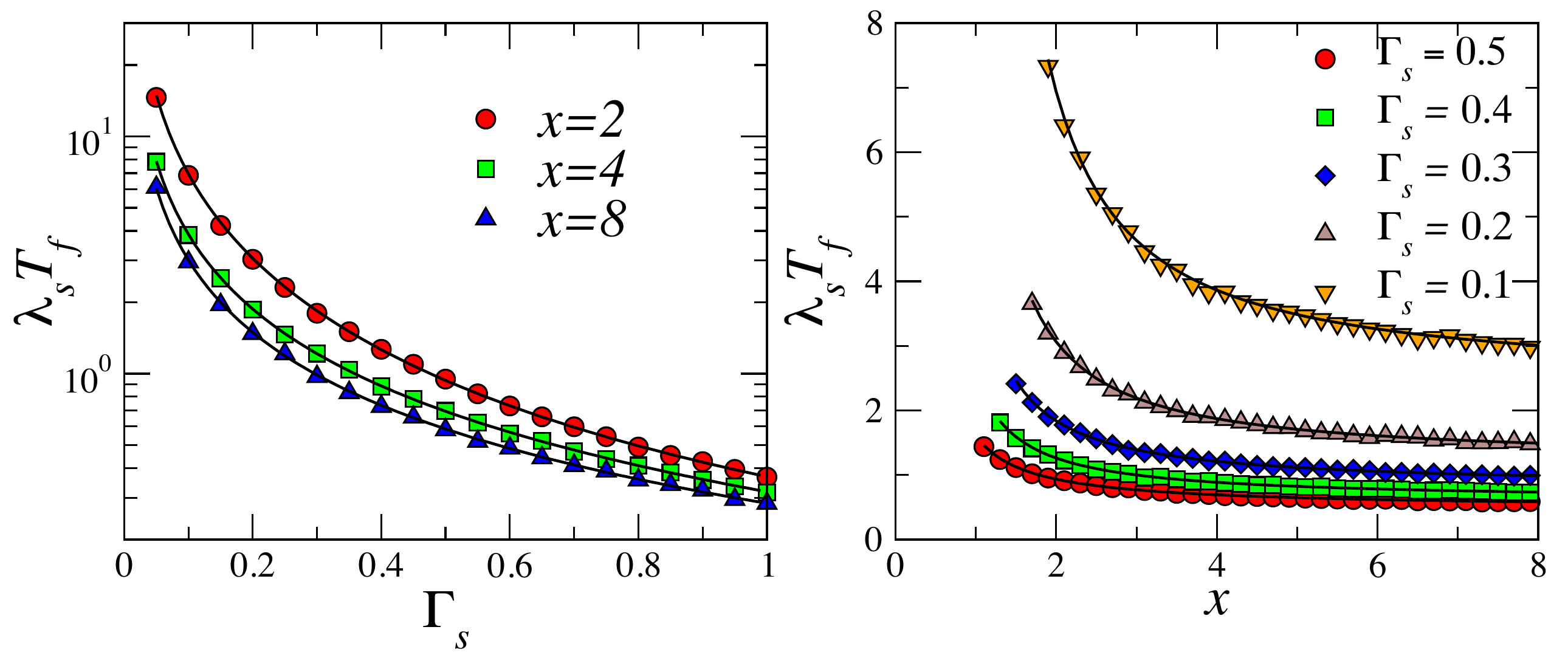}
\caption{Numerical simulations of the average crossing time for the fast group from $\Gamma_f=0$ to $\Gamma_f=1$ when $\Gamma_s$ is kept artificially fixed. The left plot shows results with a fixed value of $x$ as a function of $\Gamma_s$ and the right plot with fixed $\Gamma_s$ as a function of $x$. In both cases $\beta=\sigma=1$ and $N_s=4000$. Solid lines are obtained by numerical integration of Eq.~\eqref{mfpt}}
\label{fig:1}
\end{figure}

\subsubsection{Case $x \ge 2$}
If $x \ge 2$, the integral in Eq.~\eqref{mfpt} is bounded when $\Delta \Gamma=0$ and the behavior of $\lambda_s T_f$ is determined by the exponents $\sigma$ and $\beta$. For $\beta> \sigma^{-1}$, $\lambda_s T_{f}$ diverges in the large system size limit whereas it goes to zero whenever $\beta> \sigma^{-1}$. When $\beta = \sigma^{-1}$, $\lambda_s T_{f}$ becomes size independent and the mean crossing time depends on the value of $x$ and $\Gamma_s$. 

\begin{figure}[t]
\centering
\includegraphics[width=\columnwidth]{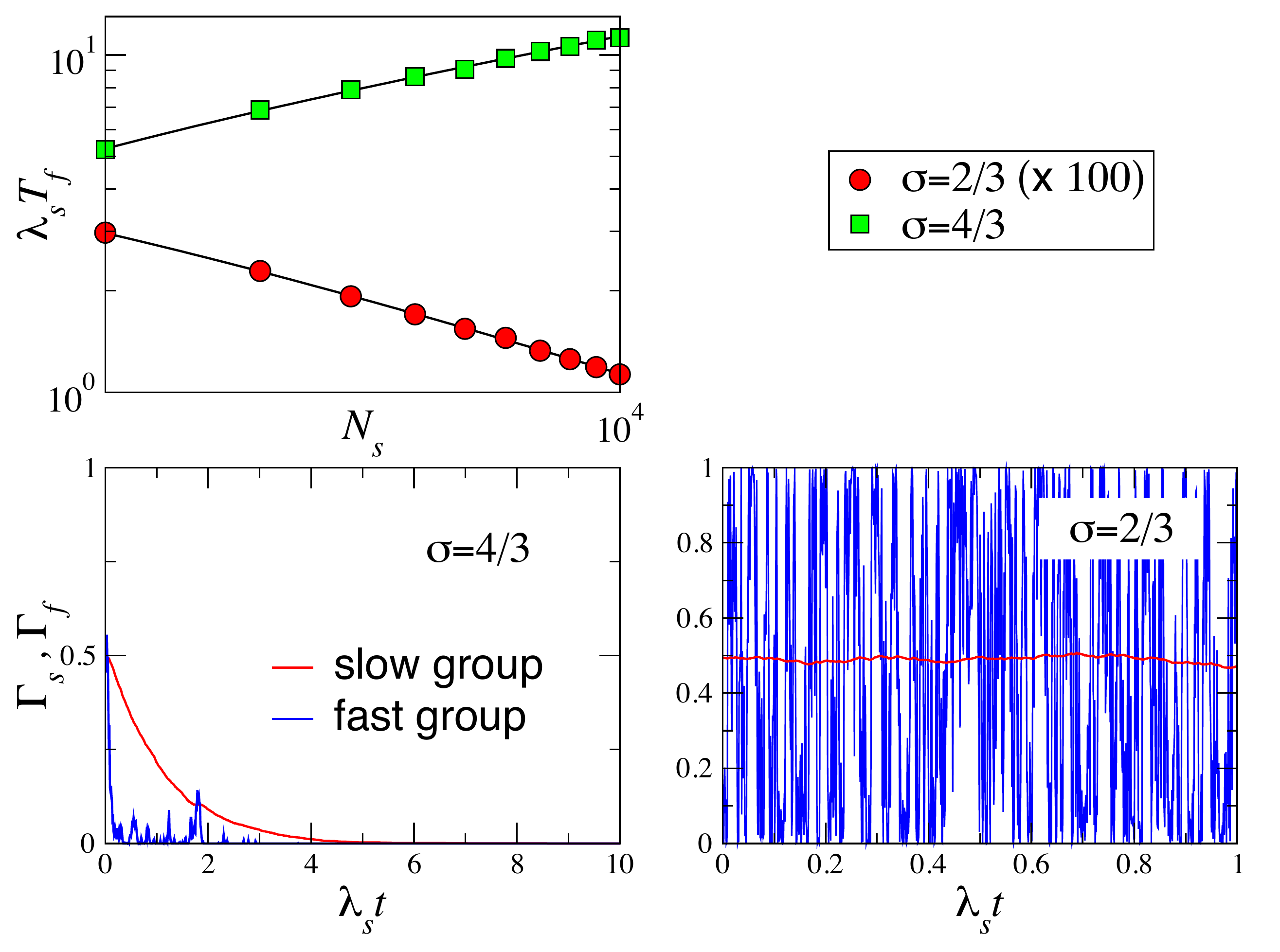}
\caption{Top: average crossing time for the fast group from $\Gamma_f=0$ to $\Gamma_f=1$ for two values of $\sigma$ as a function of the size of the slow group. In this simulations $x=2.5$, $\beta=1$, and $\Gamma_s$ is kept artificially fixed at $\Gamma_s=0.5$. Solid lines are obtained by numerical integration of Eq.~\eqref{mfpt}. Bottom: temporal evolution of the fast and slow groups for the two values of $\sigma$ considered before, starting from $\Gamma_f=\Gamma_s=0.5$ as initial conditions and $N_s=4000$.}
\label{fig:2}
\end{figure}

Figure~\ref{fig:1} shows numerical simulations with $\sigma=\beta=1$ and different values of $x$ and $\Gamma_s$ as compared to the numerical integration of Eq.~\eqref{mfpt}. The dependence of $T_f$ on $\Gamma_s$ is strong, in particular when $\Gamma_s \approx 0$. Indeed, in this case the drift term acting over fast agents induced by slow agents becomes very small making the transition very difficult. Figure~\ref{fig:2} shows the size dependence of $T_f$ for $\beta=1$, $\Gamma_s=0.5$, $x=2.5$, and two values of $\sigma$ in perfect agreement with the exact numerical solution given by Eq.~\eqref{mfpt}. Putting all the pieces together, we conclude that the global consensus time scales as
\begin{equation}
T_{con} \sim \left\{
\begin{array}{lr}
\mbox{constant} & \beta \ge 1/\sigma\\[0.5cm]
\displaystyle{\frac{\exp\{N_s^{1/\sigma-\beta}\}}{N_s^{1/\sigma-\beta}} }& \beta < 1/\sigma.
\end{array}
\right.
\end{equation}
This behavior is well illustrated in the temporal evolution of fast and slow groups shown in the bottom plots of Fig.~\ref{fig:2}.

\subsubsection{Case $1<x <2$}
In this case, the integral in Eq.~\eqref{mfpt} may diverge if $\Gamma_s<1-x/2$. Indeed, in this case, the behavior of the integrand near the upper limit makes the integral to scale as $\Delta \Gamma^{1-2(1-\Gamma_s)/x}$ and the crossing time scales as
\begin{equation}
T_{f} \sim \left\{
\begin{array}{lr}
N_s^{\beta-1/\sigma} & \Gamma_s>1-x/2\\[0.5cm]
N_s^{2\beta(1-\Gamma_s)/x-1/\sigma} & \Gamma_s<1-x/2.
\end{array}
\right.
\end{equation}
These results, however, do not change the general picture drawn in the previous case. When $\beta>1/\sigma$, $T_f$ diverges and, thus $T_{con}$ is constant. In the opposite case of $\beta < 1/\sigma$, if we start with a value of $\Gamma_s>1-x/2\beta \sigma$ then $T_f$ approaches zero and $T_{con}$ will grow exponentially fast. Finally, when $\beta < 1/\sigma$ and $\Gamma_s<1-x/2\beta \sigma$, then $T_f$ diverges and $T_{cons}$ is constant.

\section{The herding voter model with noise}

The noisy voter model has been introduced several times in different contexts during the last thirty years~\cite{LEBOWITZ1986194,Fichthorn:1989aa,Considine:1989aa,Kirman:1989,GRANOVSKY199523,Carro:2016aa}. It is a simple extension of the voter model where agents can change opinion spontaneously without any influence from their peers. Quite interestingly, this simple mechanism changes the dynamical properties of the voter model in a dramatic way. Indeed, such intrinsic noise has the effect of removing the absorbing states from the system so that, depending on the level of the intrinsic noise, the system changes from behaving in a diffusive-like fashion, like the standard voter model, or oscillating between zero and one, like in a two-states system. Quite interestingly, a similar two-states system is observed in the herding phase of the herding voter model but only for the fast group. In that case, however, such behavior is produced by the same dynamics of the herding voter model and not by any intrinsic noise decoupled from the dynamics. In this section, we merge both models and investigate the possible consequences for the global dynamics of the system.

To model the intrinsic noise, we assume that, within each agent $i$, two independent Poisson processes take place. The first one is the standard activation process of the voter model at rate $\lambda_i$, which is followed by the choice of a peer to copy her opinion. The second takes place at rate $\epsilon_i$ and it is followed by a change of the current opinion of the agent. Following~\cite{Mosquera-Donate:2015aa,Catanzaro:2005fk,Boguna:2009pi}, we can write a stochastic evolution equation for the state vector $\{n_i(t)\}$ as follows
\begin{equation}
	n_{i}(t+dt)=n_{i}(t) \left[1-\xi_{i}(t)-\phi_i(t)\right]+
	\label{eq:dinamica}
\end{equation}
\[
+\phi_i(t)\left[1-n_i(t)\right]+ \eta_{i}(t) \xi_{i}(t),
\] 
where $\xi_{i}(t)$ and $\phi_i(t)$ are random dichotomous variables that take values:
\begin{equation}
   \xi_{i}(t) = \left\{
	\begin{array}{ll}
       1 & \mbox{with probability } \lambda_{i} dt\\
       0 & \mbox{with probability } 1-\lambda_{i} dt
     \end{array}
   \right.
\end{equation}
and
\begin{equation}
   \phi_{i}(t) = \left\{
	\begin{array}{ll}
       1 & \mbox{with probability } \epsilon_{i} dt\\
       0 & \mbox{with probability } 1-\epsilon_{i} dt
     \end{array}
   \right.
\end{equation}
The stochastic process $\xi_i(t)$ controls whether node $i$ is activated during the time interval $(t,t+dt)$ whereas $\phi_i(t)$ determines whether the agent changes her opinion spontaneously~\footnote{notice that both events cannot take places simultaneously on the same time interval $(t,t+dt)$ and, thus, there is no need to introduce higher order terms.}. In the former case, the opinion of the agent is modified as
\begin{equation}
   \eta_{i}(t) = \left\{
	\begin{array}{ll}
       1 &  \mbox{with probability } \displaystyle{\sum_{j=1}^N}\frac{f(\lambda_{j})}{\sum_{i=1}^N f(\lambda_{i})} n_{j}(t) \\
       0 &  \mbox{with probability } 1-\displaystyle{\sum_{j=1}^N}\frac{f(\lambda_{j})}{\sum_{i=1}^N f(\lambda_{i})} n_{j}(t).
     \end{array}
   \right. 
\end{equation}
The first term in the right hand size of Eq.~\eqref{eq:dinamica} accounts for the case of no activity during the time interval $(t,t+dt)$, in which case the state of the agent remains the same. The second term accounts for a spontaneous change of opinion and, finally, the last term accounts for an activation of agent $i$ and the posterior adoption of the opinion of one of her peers.

In the case of a structured population with fast and slow agents, a similar analysis as the one performed in~\cite{Mosquera-Donate:2015aa} allows us to write the drift and diffusion term of fast agents as
\begin{equation}
A_f=\epsilon_f (1-2 \Gamma_f)+\alpha_{fs}(\Gamma_{s}-\Gamma_{f})
\label{eq:drift:2}
\end{equation}      
\begin{equation}
D_f=\frac{1}{N_f}\left[\epsilon_f+ \alpha_{fs}\left( \Gamma_s+\Gamma_f \left[ 1+2 \beta_{fs}-2 \Gamma_s -2 \beta_{fs} \Gamma_f\right] \right)\right].
\label{eq:diffusion:2}
\end{equation}
The equations for the slow group can be derived by switching the indices $s \leftrightarrow f$ in the previous equations.

\begin{figure}[t]
\centering
\includegraphics[width=\columnwidth]{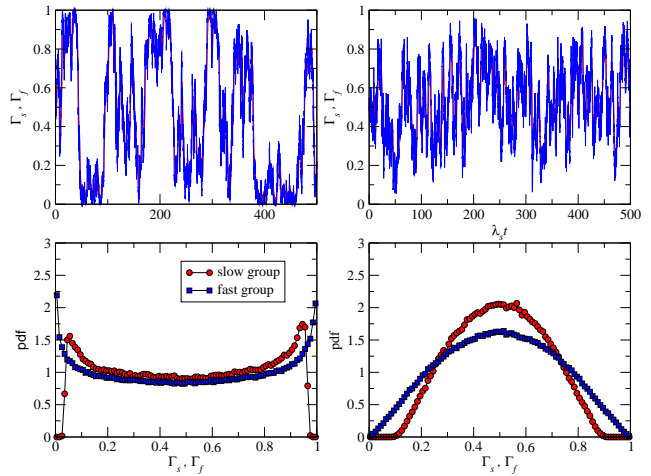}
\caption{Top: Evolution of the fast (blue) and slow (red) groups in the limit $\lambda_s t_c \gg 1$. In this simulations, we set $\lambda_s t_c=10 \sqrt{5}$, $x=1$, and $y=1.5$ (left column) and $y=0.5$ (right column). Bottom plots show the steady state probability density for both groups.}
\label{fig:3}
\end{figure}

The standard noisy voter model undergoes a phase transition between a diffusive phase and a two-states phase whenever
\begin{equation}
\epsilon <\frac{\lambda}{N}.
\end{equation}
Following this, we then define a new control parameter $y$ as
\begin{equation}
y \equiv \frac{\lambda_f}{\epsilon_f N_f}
\end{equation}
and take the limit of large system sizes by keeping $x$ and $y$ constant. This leads to the following Langevin equation for the fast group
\begin{equation}
\frac{d\Gamma_f(t)}{dt}=\frac{\lambda_f}{y N_f}[1-2 \Gamma_f]+\frac{2\lambda_f}{x N_f}[\Gamma_s-\Gamma_f]+
\label{Langevin_fast:2}
\end{equation}
\[
+\sqrt{\frac{2 \lambda_f}{N_f}\Gamma_f(1-
\Gamma_f)}\xi_f(t),
\]
whereas for the slow group we have
\begin{equation}
\frac{d\Gamma_s(t)}{dt}=\epsilon_s [1-2 \Gamma_s]+\lambda_s[\Gamma_f-\Gamma_s].
\end{equation}
Notice that Eq.~\eqref{Langevin_fast:2} defines in a natural way the characteristic time
\begin{equation}
t_c\equiv \frac{N_f}{\lambda_f}=a\lambda_s^{-1}\left(\frac{2}{x}\right)^{1/\sigma} N_s^{\beta-1/\sigma},
\end{equation}
so that by defining the dimensionless time $\tau \equiv t/t_c$, the dynamic equations for both groups read
\begin{equation}
\frac{d\Gamma_f(\tau)}{d\tau}=\frac{1}{y}[1-2 \Gamma_f]+\frac{2}{x}[\Gamma_s-\Gamma_f]+\sqrt{2\Gamma_f(1-
\Gamma_f)}\xi_f(\tau)
\label{Langevin_fast:3}
\end{equation}
and
\begin{equation}
\frac{d\Gamma_s(\tau)}{d\tau}=\frac{1}{y}[1-2\Gamma_s]+t_c \lambda_s [\Gamma_f-\Gamma_s],
\label{eq:slow}
\end{equation}
where we have assumed that the intrinsic noise is the same in both groups, that is, $\epsilon_f=\epsilon_s$. When $t_c \lambda_s \gg y^{-1} \gg 1$, the opinion of the slow group decays very fast to the opinion of the fast group so that we can approximate $\Gamma_s \approx \Gamma_f$ and the dynamics of the fast group becomes identical to the one for the standard noisy voter model. This is an extreme case of herding behavior, where the slow group behaves exactly as the fast group, whereas the fast group has an independent dynamics. Figure~\ref{fig:3} shows numerical simulations corresponding to this case for two different values of $y$, below and above the transition. As it can be clearly seen, in both cases the dynamics of the slow group is very similar to the one of the fast group, even though in one case the dynamics is diffusive-like whereas in the other it is two-states like.

\subsection{The effective potential}

In the case of $t_c \lambda_s <1$, the decay of $\Gamma_s$ is slow and we can perform an adiabatic approximation by considering that $\Gamma_s$ takes a fixed value in the Langevin equation of the fast group during the (not very large) observation time. Let us then fix $\Gamma_s$ and analyze the steady state of the fast group by using the effective potential~\cite{Gardiner:2004uv}
\begin{equation}
V_{eff}(\Gamma_f)=\ln{D_f}-2\int \frac{A_f}{D_f}d\Gamma_f.
\end{equation}
Using the expressions for the drift and diffusion terms in Eq.~\eqref{Langevin_fast:3} with $\Gamma_s$ fixed, we obtain
\begin{equation}
V_{eff}(\Gamma_f)=C(x,y,\Gamma_s) \ln{\Gamma_f}+C(x,y,1-\Gamma_s)\ln{(1-\Gamma_f)}
\label{V_eff}
\end{equation}
with 
\begin{equation}
C(x,y,\Gamma_s) \equiv \left[1-\left(\frac{1}{y}+\frac{2\Gamma_s}{x} \right) \right].
\end{equation}
As we can observe from Eq.~\eqref{V_eff}, the effective potential has logarithmic divergences both at $\Gamma_f=0$ and $\Gamma_f=1$. However, the sign of the pre-factors depends on the values of $x,y$, and $\Gamma_s$. When $y<1$, both pre-factors are negative for any value of $x$ and $\Gamma_s$. In this case, the effective potential has always a ``U'' shape and the dynamics of the fast group is diffusive-like. Of course, the actual value of $\Gamma_s$ is not constant. Nevertheless, given its slow rate of variation, we can think about the dynamics of the fast group as evolving in a slowly changing potential but that, nevertheless, does not change its qualitative properties.

\begin{figure}[t]
\centering
\includegraphics[width=\columnwidth]{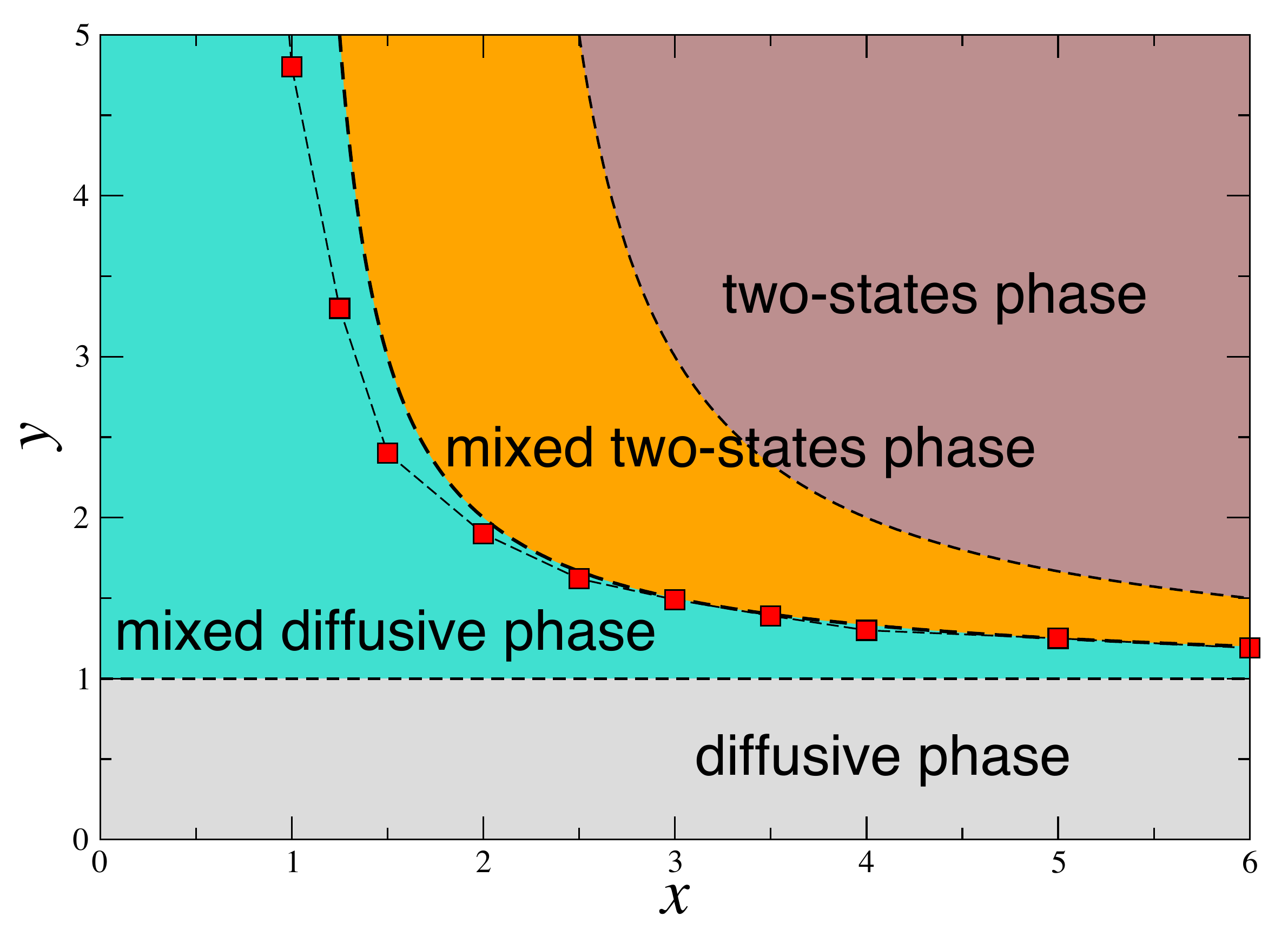}
\caption{Sketch of the different dynamical phases of the model in the parameters' space $(x,y)$ as explained in the main text. The red squares line indicates the region within the mixed diffusive phase where a two-states like dynamics for the fast group takes place.}
\label{fig:4}
\end{figure}
\begin{figure}[t]
\centering
\includegraphics[width=\columnwidth]{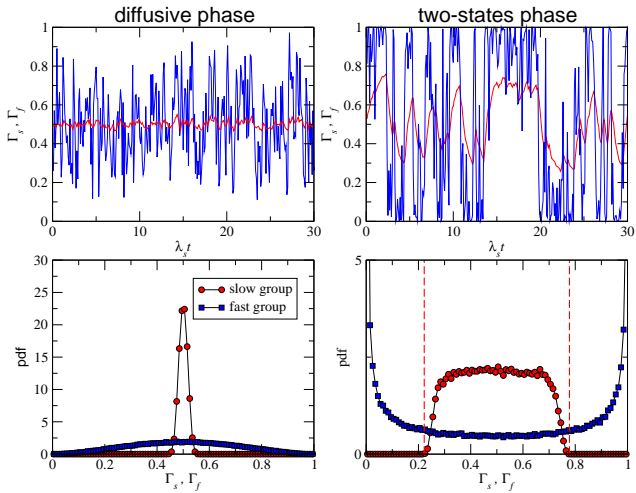}
\caption{Left: evolution of the fast (blue) and slow (red) groups in the diffusive phase with $x=1$, $y=0.5$, and $\sigma=\beta=1$, $a=0.25$, and $N_s=4000$. The bottom plot shows steady state distributions for both groups. Right: the same as in the left column but for the two-states phase with $x=3$ and $y=15$. The vertical dashed lines indicate the asymptotic value of $\Gamma_s$ when the fast group is trapped for a long time in the consensus states $\Gamma_f=0,1$, $\Gamma_s^\pm$.}
\label{fig:5}
\end{figure}

When $y>x/(x-2)$, both pre-factors are positive for any value of $\Gamma_s$. Since the boundaries $\Gamma_f=0,1$ are reflecting boundaries due to the presence of the intrinsic noise, the effective potential becomes a double-well potential with minima at the boundaries. As a consequence, the fast group will stay for a random time in one of the consensus states until it manages to jump to the other consensus state. The dynamics will then become effectively a two-states dynamics. Notice that, as in the previous case, even if $\Gamma_s$ slowly change, the qualitative shape of the effective potential remains the same. 
Between these two limit cases, we find two mixed phases. In the domain $x/(x-1)<y<x/(x-2)$, the signs of the pre-factors can be both positive --in which case, the potential has a double-well shape-- or one positive and the other negative, depending on the value of $\Gamma_s$. We call this phase a mixed two-states phase. In the domain $1<y<x/(x-1)$, the signs of the pre-factors can be both negative --in which case, the potential has a ``U'' shape-- or one positive and the other negative, depending on the value of $\Gamma_s$. We call this phase a mixed diffusion phase. All these phases are shown in Fig.~\ref{fig:4}.

Figure~\ref{fig:5} shows numerical simulations of the diffusive and two-states phases, respectively. As expected, in the diffusive phase, both groups fluctuate symmetrically around 1/2, although the fast group does it with higher fluctuations. In the two-states phase, the fast group oscillates between the two consensus states, as also expected. At the same time, the slow group tries to catch up following the current state of the fast group, until it reaches a steady value, which correspond to the stationary solution of Eq.~\eqref{eq:slow} when $\Gamma_f=0$ or $1$, that is,
\begin{equation}
\Gamma_s^{\pm}=\frac{1}{2}\left[ 1\pm \frac{y t_c \lambda_s}{2+y t_c \lambda_s} \right].
\end{equation}
These values are indicated by the dashed vertical lines in the corresponding histograms.

The behavior of the system within the mixed two-states phase is qualitatively similar to the one in the two-states phase. The main difference arises for high values of the term $y t_c \lambda_s$, so that $\Gamma_s^{\pm} \approx (1\pm1)/2$. In this case, if we start the dynamics with $\Gamma_s=0.5$, the effective potential has initially a double-well shape and the fast group will fall in one of the consensus states. The slow group will then start approaching its ``steady'' configuration $\Gamma_s^{\pm}$ and, eventually, the effective potential will change its qualitative shape to become a slope, trapping the fast group in the current consensus state with more intensity. Eventually, the fast group will manage to scape from this state -- modifying the shape of the effective potential-- and get trapped in the other consensus state, so behaving again as a two states-system. However, this process will take more time as compared to the two-states phase, where the potential does not change its qualitative shape. In turn, this implies that the slow group will spend more time near $\Gamma_s^\pm$ and, thus, the steady fluctuations of the slow group will be higher.

\begin{figure}[t]
\centering
\includegraphics[width=\columnwidth]{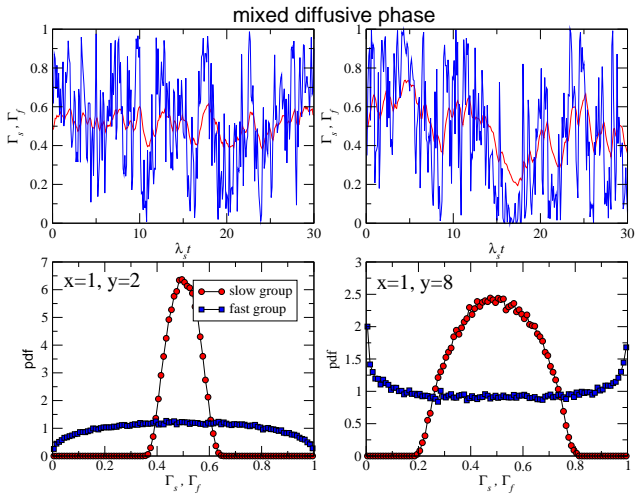}
\caption{Left: evolution of the fast (blue) and slow (red) groups in the mixed diffusive phase with $x=1$, $y=2$, and $\sigma=\beta=1$, $a=0.25$, and $N_s=4000$. The bottom plot shows steady state distributions for both groups. Right: the same as in the left column but for $x=1$ and $y=8$.}
\label{fig:6}
\end{figure}

Finally, the behavior of the system in the mixed diffusive phase can be different depending on the value of $x$ and $y$. For a fixed value of $x$ and low values of $y$, the fluctuations of both groups are small, the effective potential will never change its qualitative shape, and the system has a diffusive-like behavior. However, for higher values of $y$ fluctuations are important enough to take $\Gamma_s$ to the point where the effective potential changes from having a ``U'' shape to a slope shape. When this event takes place, the fast group is pushed to the corresponding consensus state and remains there until fluctuations takes it to the other consensus state. The system thus behaves effectively as a two-states system. Figure~\ref{fig:6} shows numerical simulations of the mixed diffusive phase showing these phenomena whereas Fig.~\ref{fig:4} shows the empirical line in the plane $(x,y)$ where such behavior occurs.

\section{Conclusions}

As we have seen, the addition of small variations to the classical voter model increases the range of possible dynamical behaviors dramatically. Heterogeneity in the activity rates of agents, coupled with a preference choice for active agents, induce the emergence of collective leadership in a fraction of the population while the rest simply follow the opinion of the leading group. This has important consequences for the global consensus time, which now range from being a constant value independent of the system size to an exponential function of the system size, in stark contrast to the standard voter model. On the other hand, the addition of intrinsic noise to the previous model makes its dynamics even richer, with the emergence of four well resolved dynamical phases with distinct behavior separations. Speculatively, it might be possible to attribute these phases to observable modes of social behavior in large groups. E.g. sudden jumps of the average opinion to one of the consensus states can be interpreted as informational cascades, where a plurality of agents at the same time change their attitudes in one direction; addition of spontaneous opinion changes makes the group more tolerant to polarized opinion oscillations --- requiring more ``fast'' agents to effect the opinion of the ``slow'' part of the group and move the entire group to one of the polar extremes. We hope that, despite the simplicity of the model, our results will increase our understanding of the opinion dynamics of large groups of interacting agents in fields such as economy or sociology.

\begin{acknowledgments}
This work was supported by: the European Commission within the Marie Curie ITN ``iSocial'' grant no.\ PITN-GA-2012-316808; a James S. McDonnell Foundation Scholar Award in Complex Systems; the ICREA Academia prize, funded by the {\it Generalitat de Catalunya}; the MINECO projects nos.\ FIS2010-21781-C02-02 and FIS2013-47282-C2-1-P (AEI/FEDER, UE);  and the {\it Generalitat de Catalunya} grant no.\ 2014SGR608.
\end{acknowledgments}


%

\end{document}